\pdfoutput=1 
\documentclass{JINST}
\let\ifpdf\relax

\title{Performance of Ultra-Fast Silicon Detectors}

\author{N. Cartiglia$^a$\thanks{Corresponding author.}~,
   M. Baselga$^b$$^c$, G. Dellacasa$^a$, S.Ely$^b$$^d$, V. Fadeyev$^b$, Z. Galloway$^b$, S. Garbolino$^a$, F. Marchetto$^a$, S. Martoiu$^a$,  G. Mazza$^a$, J. Ngo$^b$, M. Obertino$^a$$^e$, C. Parker$^b$, A. Rivetti$^a$, D. Shumacher$^b$, H.F-W.Sadrozinski$^b$,A. Seiden$^b$, A. Zatserklyaniy$^b$\\
\llap{$^a$}INFN Torino,\\
  Via Pietro Giuria 1, Torino, Italia\\
E-mail: \email{cartiglia@to.infn.it} \\
\llap{$^b$}Santa Crus Institute for Particle Physics\\ UC Santa Cruz, CA, 95064, USA.\\
\llap{$^c$}Permanent address: Centro Nacional de Microelectronica, CNM-IMB (CSIC), Barcellona, Spain.\\
\llap{$^d$}Permanent address: Dept. of Physics, Syracuse, Univ., Syracuse, NY 13210, USA. \\
\llap{$^e$} Universit\`a del Piemonte Orientale, Novara, Italia.\\
}

\abstract{The development of Low-Gain Avalanche Detectors  has opened up the possibility of manufacturing silicon detectors with signal larger than that of traditional sensors. In this paper we explore the timing performance of Low-Gain Avalanche Detectors, and in particular we demonstrate the possibility of obtaining ultra-fast silicon detector  with time resolution of less than 20 picosecond. }

\keywords{detector; fast; silicon, timing}

\begin{document}

\section{Introduction}\label{sec:intro}

The possibility to use and control charge multiplication in un-irradiated silicon detectors has been the subject of intense study within the RD50 collaboration~\cite{RD50}. The basic mechanism to obtain charge multiplication  is to create, within the bulk of a silicon sensor,  a large volume  where the electric field is high enough so that the drifting electrons will generate a controlled, low gain avalanche. Charge multiplication in silicon detector follows, for a constant field,  a typical exponential behaviour:

\begin{equation}
N(x) = N_o * e^{\alpha*x} = G * N_o,
\end{equation}

where at a field $ V = 270~kV/cm$ the value of $\alpha$ for electrons is $\alpha_e \sim 0.7$ pair/$\mu$m while for holes is $\alpha_h \sim 0.1$ pair/$\mu$m.
\\

\begin{figure}[tb]
\centering
\includegraphics[width=.4\textwidth]{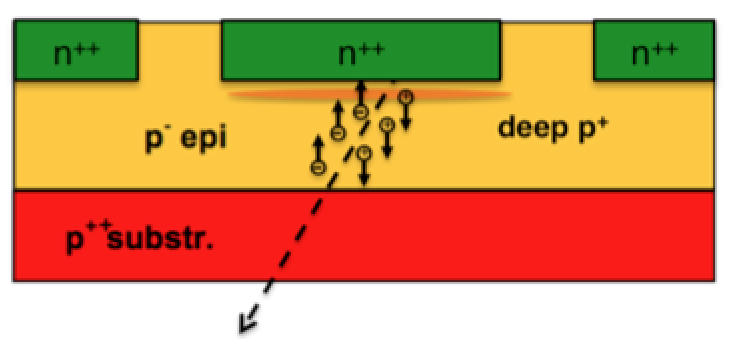}
\caption{Schematic of a Low Gain Avalanche Diode. The extra deep $p^+$ layer creates a strong electric field that generates charge multiplication.}
\label{fig:sk}
\end{figure}

Low-gain avalanche detectors (LGAD), as develop by CNM ~\cite{CNM1, CNM2}, are $n-on-p$ silicon sensors with a high ohmic $p$ bulk which have a $p^+$ implant extending several microns underneath the n-implant. Figure~\ref{fig:sk} shows the $n^{++}-p^+-p-n^{++}$ structure of a LGAD.

This implant generates a large local field at a depth of about $1-5 ~\mu$m. The doping concentration of the $p^+$  implant is chosen to generate a gain of 10-100, in contrast to a gain of $10^4$ or more in silicon photomultipliers (SiPM) and multi-pixel photon counters (MPPC). LGAD sensors work by inducing multiplication for electrons, while the hole multiplication, given the field and depth values involved, is insignificant. Therefore, LGAD sensors do not have a positive feedback loop formed by the concurrent electron and hole multiplication processes, present in SiPM, which causes dead time after the  avalanche. 
\\

\subsection{Ultra-Fast Silicon Detectors}
\label{sec:ufsd}

The design of ultra-fast silicon detectors (UFSD)~\cite{HFWS1, HFWS2, HFWS3} exploits the effect of charge multiplication in LGAD to obtain a silicon detector that can concurrently measure with high accuracy time and space. The development of UFSD  will open up a range of new opportunities for applications that benefit from the combination of position and timing information. UFSD are the first detectors able to perform 4-dimensional tracking of charged particles with a very good space and time resolution: $\sigma_t \sim 10-30\; ps, \; \; \sigma_x \sim 20-50 \mu$m. 

In its foreseen design, UFSD employs a dedicated ASIC chip for the read-out, in an hybrid configuration. The pixel size of UFSD needs to be large enough to house the necessary electronic circuits: currently we foreseen a minimum pixel size of 50-100 $\mu$m, depending on the technology used for the ASIC design. 
\\

In the following we review the basic ingredients of a time-tagging detector and the state of the art of silicon detector timing. We propose a general parametrization of the timing characteristics of a detector and we use it to predict the timing performances of UFSD.

\section{Time-Tagging detectors }\label{sec:tag}

\begin{figure}[tb]
\centering
\includegraphics[width=.8\textwidth]{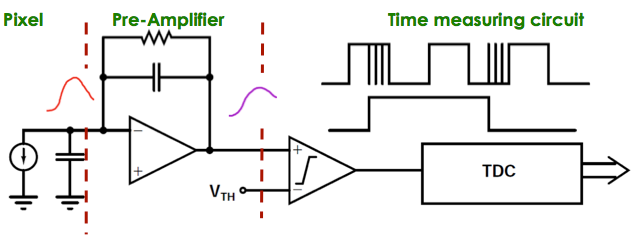}
\caption{Main components of a time-tagging detectors. The time is measured when the signal crosses the threshold.}
\label{fig:tag}
\end{figure}

Figure~\ref{fig:tag} shows the main components of a time-tagging detector. For a review of current trends in electronics see for example~\cite{Rivetti}. The silicon sensor, a pixel in the picture, is read-out by a pre-amplifier that shapes the signal. The shaper's output is then compared to a fixed threshold to determine the time of arrival. In the following we will use this simplified model to explore the UFSD timing capabilities, while we will not consider more complex and space-consuming approaches such as waveform sampling.
\\

The time resolution $\sigma_t$ can be expressed as the sum of three terms: (i) Time Walk , (ii) Jitter, and (iii) TDC binning:
\begin{equation}
\label{eq:main}
\sigma_t^2 = \sigma_{TW}^2 + \sigma_{J}^2+\sigma^2_{TDC}.
\end{equation}

TDC binning introduces a fix uncertainty equal to $\sigma_{TDC} = TDC_{bin}/\sqrt{12}$. As the performance  of TDCs become faster and faster~\cite{Rivetti}, we assume $TDC_{bin} = $ 20 ps and therefore this effect will not be important.

\subsection{Time Walk}\label{subsec:tw}

\begin{figure}[tb]
\centering
\includegraphics[width=.8\textwidth]{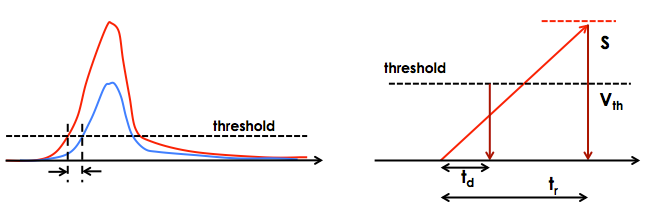}
\caption{Left pane: Signals of different amplitude cross a fix threshold at different times, generating a delay on the   on the firing of the discriminator  that depends upon the signal amplitude. Right pane: a linear signal, with amplitude $S$ and rise time $t_r$  crosses the threshold $V_{th}$ with a delay $t_d$.}
\label{fig:TW}
\end{figure}

The term \emph{Time Walk} indicates the unavoidable effect that larger signals  cross a given threshold earlier than smaller ones, Figure~\ref{fig:TW}, left pane.

Let's assume for simplicity a linear signal, with amplitude $S$ and rise time $t_r$. This signal crosses the threshold $V_{th}$ with a delay $t_d$, Figure~\ref{fig:TW}, right pane. From  the following identity $ t_d : t_r = V_{th}:S$ we can derive:
\begin{equation}
t_d = \frac{t_rV_{th}}{S}.
\end{equation}

Setting the value of the threshold to $V_{th} = S_o/3$, where $S_o$ is the most probable value of the signal amplitude, and assuming a signal variation of $S_o/3 < S < 5*S$, then the delay can vary from $t_d = t_r$ for $S = S_o/3$  to  $t_d = t_r/15$ for $S = 5S_o$. 

We define the timing uncertainty due to time walk as the rms of the delay time distribution:
\begin{equation}
\sigma_t = [t_d]_{RMS} = [\frac{t_rV_{th}}{S}]_{RMS}.
\end{equation}
As it's clear from this expression, to minimize the effect of time walk we need to use the lowest possible value of the threshold. In the following we will use $V_{th} = 10*N$, where $N$ is the noise measured at the pre-amplifier output. 

\begin{figure}[tb]
\centering
\includegraphics[width=.6\textwidth]{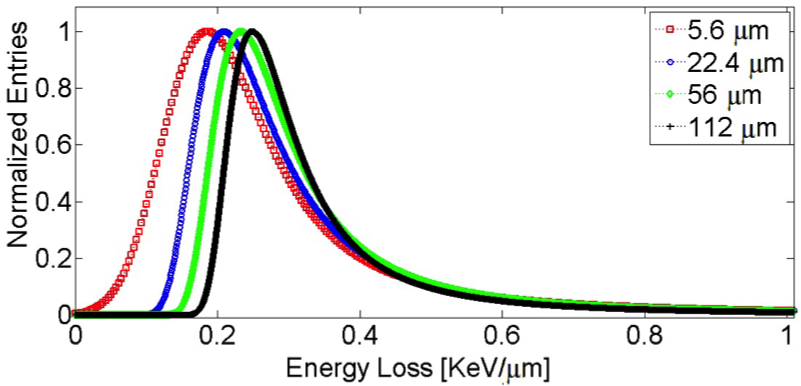}
\caption{Average energy loss per micron of a ionizing particle in silicon layers of different thicknesses.}
\label{fig:Me}
\end{figure}

In silicon detectors, the amplitude $S$ varies according to a Landau distribution. According to~\cite{meroli}, the average energy loss per micron in the bulk of silicon  decreases for thinner detector while the Landau width increases, Figure~\ref{fig:Me}. The Landau Most Probable Value ($mpv$) and width $\Delta S$ for a detector of thickness $d$ are given by:
\begin{eqnarray}
\label{eq:mpv}
{\rm mpv = 0.027*ln(d) + 0.126} \\
\label{eq:da}
{\rm \Delta S/S = 0.7079 *d^{-0.266}}.
\end{eqnarray}

Thin sensors suffer therefore of two additional problems  with respect of thicker sensors: their average energy loss per micron is smaller and the variations are larger. Both these effects cause an enhanced time walk. 

\begin{figure}[tb]
\centering
\includegraphics[width=.6\textwidth]{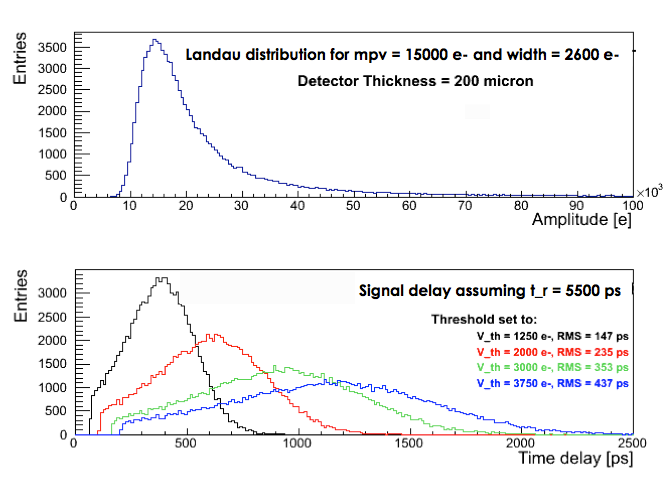}
\caption{Top pane: Landau distribution of the energy lost by a minimum ionizing particle in a 200 $\mu$m thick silicon detector. Bottom pane: delay distributions for four different values of the threshold.}
\label{fig:TDelay}
\end{figure}

Using equations (\ref{eq:mpv}) and (\ref{eq:da}), it is possible to generate the appropriate Landau distribution for any given detector thickness. Figure~\ref{fig:TDelay} (top pane) shows the case of $d = 200 \mu m$, while 
Figure~\ref{fig:TDelay} (bottom pane) shows the delay distributions for four different values of the threshold $V_{th}$, together with their RMS. The value of the time jitter for a shaping time of $t_r = 5500$ ps is therefore $\sigma_{TW} \sim 200-500 $ ps, depending upon the chosen value of  the threshold.

\subsubsection{Time Walk Mitigation Techniques}
The effect of time walk on the time determination can be greatly reduced if the signal amplitude is known. In this case, a correction function can be easily implemented. For this purpose, two techniques are commonly used: Time-over-Threshold (ToT) and Constant Fraction Discriminator (CDF). In this article we will not examine the details of these two techniques, however we will assume, very conservatively, that a factor of 3 in reduction of $\sigma_{TW}$ can be achieved due to their use.

\subsection{Jitter}
\begin{figure}[tbp] 
\centering
\includegraphics[width=.6\textwidth]{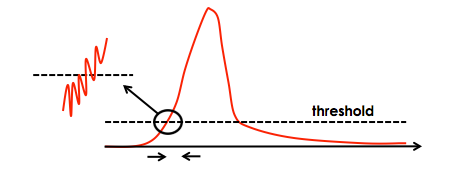}
\caption{Noise causes an uncertainty on the time when the  comparator fires.}
\label{fig:jitter}
\end{figure}

Time uncertainty caused by the early or late firing of the comparator due to the presence of noise on the signal is called \emph{jitter}. Figure~\ref{fig:jitter} shows this effect.

Jitter is directly proportional to the noise $N$ of the system and it is inversely proportional to the slope of the signal around the value of the comparator threshold. Assuming a constant slope we can write $dV/dt = S/t_r$ and therefore:
\begin{equation}
\sigma_J = \frac{N}{dV/dt} = \frac{t_r}{S/N}.
\end{equation}

\section{A Parametrization of $\sigma_t$}
Using the explicit expressions of $\sigma_{TW},\; \sigma_{J}$ and $\sigma_{TDC}$, equation~\ref{eq:main} can be rewritten as:

\begin{equation}
\label{eq:maine}
\sigma_t^2 = ([\frac{t_r V_{th}}{S}]_{RMS})^2 + (\frac{t_r}{S/N})^2+(\frac{TDC_{bin}}{\sqrt{12}})^2.
\end{equation}
Let's introduce the following quantities:
\begin{itemize}
\item[\bf{d}:] Detector thickness [micron]
\item[\bf{l}:] Pixel pitch (assuming square pixels) [micron]
\item[\bf{$C_{Det}$}:] Detector capacitance: $C_{Det} = \epsilon\epsilon_o\frac{l*l}{d}+0.2*4l+50$ fF. \\
The first term accounts for the capacitance to the back-plane, the second for the contribution from the neighbours and the third one for constant stray contributions.
\item[\bf{N}:] Noise: $N \propto \frac{C_{Det}}{\sqrt{t_r}}$ \\
 We assume that is dominated by the voltage term. 
\item[\bf{S}:] Signal. The signal amplitude is determined by the detector thickness via equations~\ref{eq:mpv} and \ref{eq:da}.
\item[\bf{$t_r$}:] Preamplifier rise time. 
\item[\bf{$V_{th}$}:] Comparator threshold. Set to 10 times the noise level:  $V_{th} = 10 * N$
\item[\bf{$TDC_{bin}$:}] TDC bin width. We consider a value of 20 ps.
\end{itemize}

\subsection{Choice of Preamplifier Rise Time}
Several  effects such as the system noise $N$, the collected charge and the possibility to generate fake hits on neighbouring sensors should be considered when deciding the preamplifier rise time.  In particular, the shaping time should be compared to the charge collection time ($t_{col}$) in the sensor to evaluated the amount of charges collected within $t_r$: the signal $S$ increases until $t_r \sim t_{col}$ while it's a constant  for $t_r > t_{col}$. Table~\ref{tab:time} shows explicitly the dependence of several factors of equation~\ref{eq:maine} upon $t_r$.

\begin{table}[tbp]
\caption{Dependence of various terms upon the preamplifier shaping time}
\label{tab:time}
\smallskip
\centering
\begin{tabular}{|c|c|}
\hline
Noise & $ C_{Det}/\sqrt{t_r}$ \\ \hline
Signal &  $t_r < t_{col} \rightarrow S\propto t_r$ \\ 
& $t_r > t_{col} \rightarrow S\propto Const $ \\ \hline
Threshold & $ \propto N \propto C_{Det}/\sqrt{t_r}$. \\
\hline
\end{tabular}
\end{table}

\begin{figure}[tb]
\centering
\includegraphics[width=.6\textwidth]{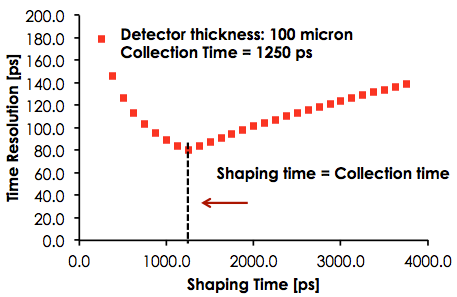}
\caption{Time resolution  $\sigma_t$ for an $ l = 100 \; \mu$m  pixel as a function of the amplifier shaping time $t_r$.}
\label{fig:coll}
\end{figure}

Using the expressions of Table~\ref{tab:time} in equation~\ref{eq:maine}, we can derive the dependence of $\sigma_t$ upon $t_r$:
\begin{eqnarray}
 t_r < t_{col} \rightarrow \sigma_t \propto \frac{C_{det}}{\sqrt{t_r}} \\
 t_r > t_{col} \rightarrow \sigma_t \propto C_{det} * \sqrt{t_r}.
\end{eqnarray}
Assuming a detector thickness of 100 micron, with a collection time $t_{col} = 1250 $ ps, Figure~\ref{fig:coll} shows the time resolution  $\sigma_t$ for an $l = 100 \;\mu$m  pixel as a function of the amplifier shaping time $t_r$.

Time resolution is therefore minimized for $t_r \sim t_{col}$, relation that will be always used in the remaining part of this paper. 

\section{Results}

\begin{figure}[tb]
\centering
\includegraphics[width=.8\textwidth]{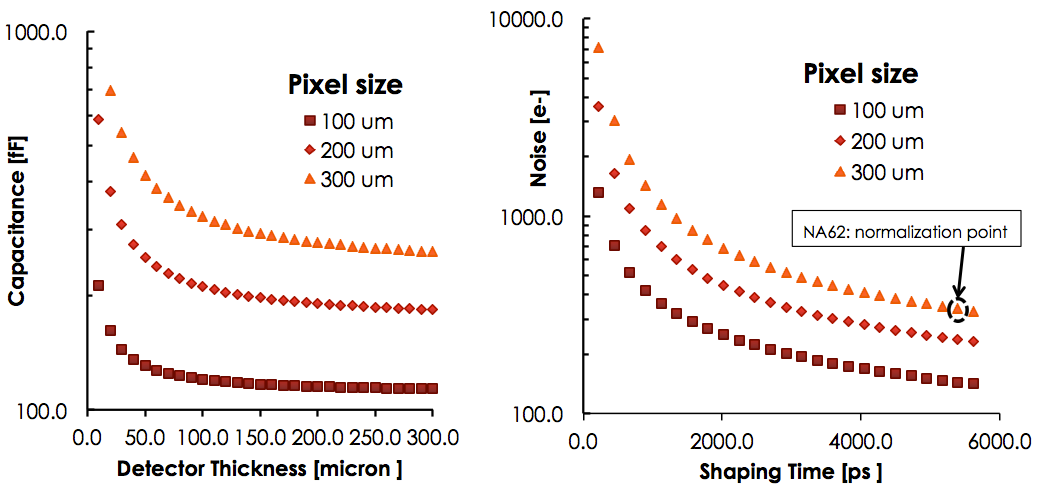}
\caption{Interplay among key parameters of the parametrization of $\sigma_t$.}
\label{fig:par}
\end{figure}

The interplay among key parameters of  equation~\ref{eq:maine} is shown in Figure~\ref{fig:par}. The detector thickness $d$ and pixel size $l$ determine the capacitance $C_{det}$ and shaping time $ t_r$, which in turn determines, together with $C_{det}$ the noise $N$. The curves have been normalized to the existing NA62 Gigatracker system~\cite{NA62}: ${\rm d = 200 \; \mu m; \;  l = 300\; \mu m; \;  t_r = 5500\; ps \; and \; N = 300\; e^-}$, accounting for the fact that the Gigatracker shaping time it's longer  than the collection time. 

\subsection{State of the Art}

\begin{figure}[tb]
\centering
\includegraphics[width=1.\textwidth]{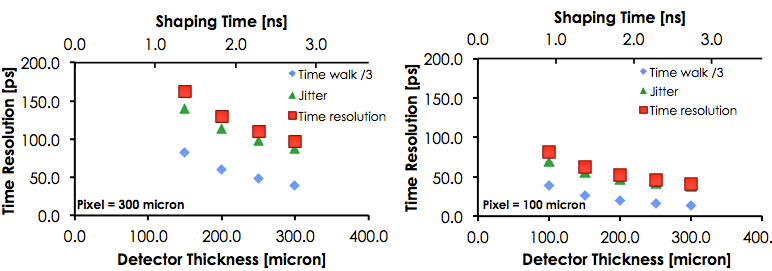}
\caption{The left (right) pane shows $\sigma_t$, and its two parts $\sigma_{TW}$ and $\sigma_J$, for a 300 $\mu$m ( 100 $\mu$m) pixel sensor as a function of detector thickness.}
\label{fig:soa}
\end{figure}

With the assumptions outlined above, the state of the art of timing capability in silicon sensors is shown in Figure~\ref{fig:soa}. The left (right) pane shows $\sigma_t$, and its two parts $\sigma_{TW}$ and $\sigma_J$, for a $l = 300 \; \mu$m ($l = 100 \;\mu$m) pixel sensor as a function of detector thickness. The contribution from time walk has been reduced by a factor of three, considering the effect of a ToT or CFD circuit. The secondary x-axis at the top of each plot shows the appropriate shaping time. 

Our parametrization shows that the best time resolution is obtained for thicker sensors, driven by higher signals,  while larger pixel size have worse time resolution due to their higher capacitance. We find $\sigma_t \sim 100$ ps for a $l = 300 \; \mu$m pixel sensor of 250-300 $\mu$m thickness, while, for the same sensor thickness but a pixel size of $l = 100 \; \mu$m we obtain  $\sigma_t \sim 50$ ps.

\subsection{Ultra-Fast Silicon Detector}

\begin{figure}[tb]
\centering
\includegraphics[width=.6\textwidth]{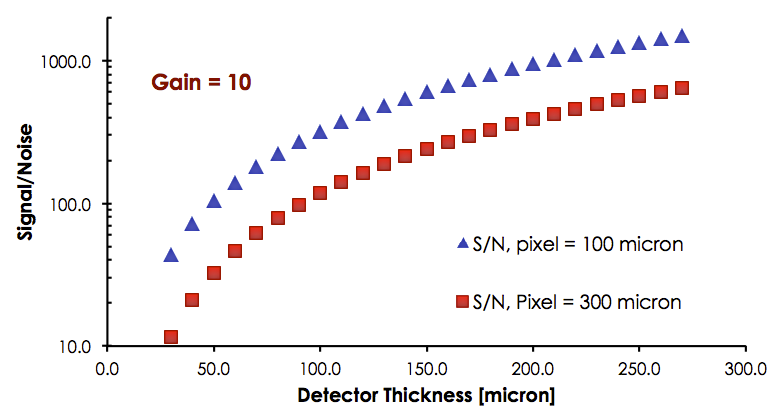}
\caption{ $S/N$ ratio for a $l = 100 \;\mu$m and a $l = 300 \; \mu$m pixel detector as a function of the detector thickness for an UFSD with gain = 10.}
\label{fig:noise}
\end{figure}

As equation~\ref{eq:maine} shows, the key to better time resolution is the possibility to increase the $S/N$ ratio.  Exploiting the larger signal from LGAD, we foresee the possibility to build ultra-fast silicon detectors. In the following we will assume that the $S/N$ ratio in UFSD increases by a factor of 10, obtained by means of a larger signal without an equivalent increase of the noise. 
Figure~\ref{fig:noise} shows the $S/N$ ratio for a $l = 100 \; \mu$m and a $l = 300 \; \mu$m pixel detector as a function of the sensor thickness for an UFSD with gain = 10. 

It's important to notice that charge collection time $t_{col}$ for UFSD is actually longer than that in traditional pixel detectors, as it comprises of the usual time drift of the charges towards the respectively electrodes, plus the time taken by the holes produced in the the multiplication layer to drift back to the $p^{++}$ electrode. 

\begin{figure}[tb]
\centering
\includegraphics[width=1.\textwidth]{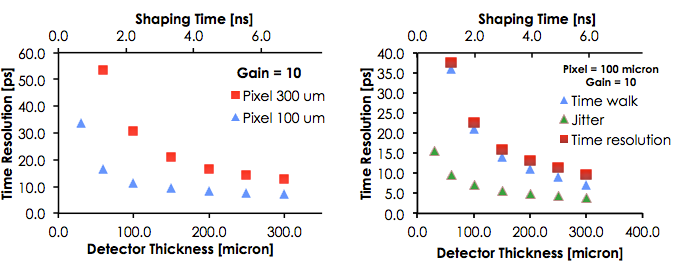}
\caption{Time resolution $\sigma_t$ for two different pixel sizes ($l = 100 \; \mu$m and $ l = 300 \;\mu$m) as a function of the detector thickness for a UFSD with Gain = 10.Right pane: time resolution $\sigma_t$ for a pixel sizes of $l = 100 \;\mu$m without any time walk mitigation circuit, as a function of the detector thickness for a UFSD with Gain = 10.}
\label{fig:gain10}
\end{figure}

The effect of the increased $S/N$ ratio in UFSD is visible in Figure~\ref{fig:gain10}(left pane), where a time resolution $\sigma_t < 20 $ ps is achieved for detector thicknesses above $d = 50 \; \mu$m for a $l = 100 \;\mu$m pixel, while a $l = 300 \; \mu$m pixel detector reaches an analogous time resolution for thicknesses above $ d = 150 \; \mu$m.

A very large $S/N$ is also a great benefit for time-walk correction: as the signal $S$ increases while the threshold $V_{th}$ does not change,  the time-walk becomes quite small. The time resolution $\sigma_t$ without any time-walk mitigation circuit is shown in Figure~\ref{fig:gain10} (right pane) for a $l = 100 \; \mu$m pixel size:  a time resolution of less that 20 ps is achievable for detector thicknesses above 100 $\mu$m. The possibility to have good time resolution without time walk correction greatly simplifies the associated electronics, allowing for smaller pixels and lower power consumption. 


\subsubsection{Additional Sources of Timing Errors}
Time Walk  addresses the impact of amplitude variations on $\sigma_t$ while the jitter term parametrizes the effect of system noise. There are also additional effects that might contribute to $\sigma_t$ by modifing the shape of the signal, for example track location and direction, variation of ionization location along the track, $\delta$ rays and diffusion~\cite{Parker}. These effects are, however, minimized by the geometry of planar pixel sensors as  the Ramo weigthing field  is, contrary to 3D detectors,  almost constant within the bulk and the drift velocity is always saturated.

\section{Summary}

\begin{figure}[tb]
\centering
\includegraphics[width=.6\textwidth]{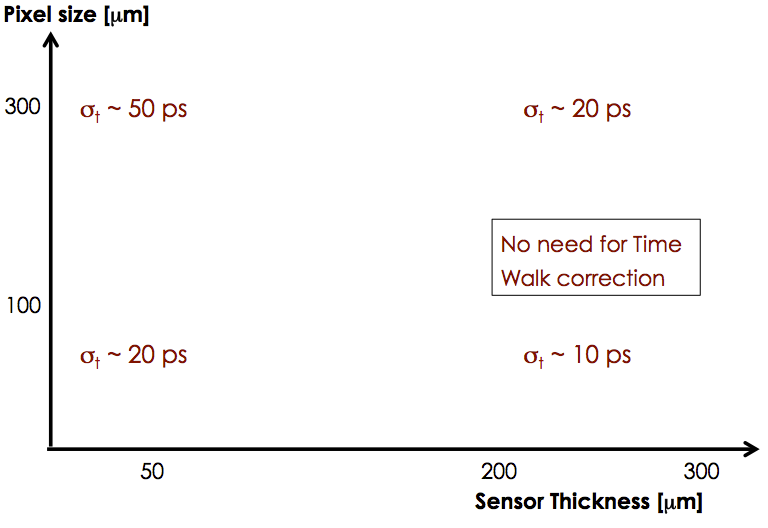}
\caption{Summary of the results obtained with our parametrization: thicker detectors have better time resolution, at a price of higher occupancy, as the drift time is longer. Smaller size pixels, due to the lower capacitance value, offer better performances. The combination of small size pixels ($ l < 150 \;\mu$m) and thick detector ($d \sim 200 - 300 \; \mu$m) allows for a simplified electronics, as the time-walk compensating circuit might not be necessary. }
\label{fig:summary}
\end{figure}

We propose a parametrization to evaluate the timing performance of Ultra-Fast Silicon Detectors, based on the LGAD concept. The increased $S/N$ ratio is the key for the excellent timing performance. Figure~\ref{fig:summary} presents our findings: thicker detectors have better time resolution, at a price of higher occupancy, as the drift time is longer. Smaller size pixels, due to the lower capacitance value, offer better performances. The combination of small size pixels ($ l < 150 \;\mu$m) and thick detector ($d \sim 200 - 300 \; \mu$m) allows for a simplified electronics, as the time-walk compensating circuit might not be necessary.

\acknowledgments

The authors wish to thank Gian-Franco Dalla Betta and Gregor Kramberg  for their insights and suggestions.

\end{document}